\documentclass[useAMS,usenatbib,fleqn]{mnras}

\usepackage{rotating}
\usepackage{longtable}
\usepackage{graphicx}
\usepackage{aas_macros}
\usepackage[T1]{fontenc}
\usepackage{ae,aecompl}
\usepackage{newtxtext,newtxmath}

\newcommand{\teff}{$T_{\rm eff}$}

\def\lgg{log\,${g}$}

\newcommand{\vs}{$v_{\rm e}\sin i$}
\newcommand{\vr}{$V_{\rm r}$}

\newcommand{\kms}{km\,s$^{-1}$}
\def\bs{$\langle B_{\rm s} \rangle$}

\def\ione{\,{\sc i}}
\def\ii{\,{\sc ii}}
\def\iii{\,{\sc iii}}
\def\abun{$\log(N_{el}/N_{tot})$}
\newcommand{\bmag}{\textsc{BinMag6}}

\title[Evolutionary status of HD~188041, HD~111133, \& HD~204411]
{Fundamental parameters and evolutionary status of the magnetic chemically 
	peculiar stars HD~188041 (V1291 Aquilae), HD~111133 (EP Virginis), and HD~204411. Spectroscopy versus interferometry.
	\thanks{Based on observations made with UVES spectrograph of the ESO VLT telescope under program ID 68.D-0254, with SARG spectrograph attached to TNG telescope, and with the VEGA/CHARA spectro-interferometer.}
}
\author[A. Romanovskaya et al.] {\parbox{\textwidth}{A. Romanovskaya$^{1}$\thanks{E-mail:
			annarom@inasan.ru}, T. Ryabchikova$^{1}$, D. Shulyak$^{2}$, K. Perraut$^{3}$, G.~Valyavin$^{4,5}$, T.~Burlakova$^4$, G.~Galazutdinov$^{6,7}$
	}\\ 
	\\
	$^{1}$Institute of Astronomy, Russian Academy of Sciences, Pyatnitskaya 48, 119017, Moscow, Russia\\
	$^{2}$Max-Planck Institute f\:ur Sonnensystemforschung, Justus-von-Liebig-Weg 3, D-37077, G\:ottingen, Germany\\
	$^{3}$University Grenoble Alpes, IPAG, 38000, Grenoble, France\\
	$^{4}$Special Astrophysical Observatory, Russian Academy of Sciences, Nizhnii Arkhyz 369167, Russia\\
	$^{5}$Federal State Budget Scientific Institution ``Crimean Astrophysical Observatory of RAS'', Nauchny, 298409, Crimea, Russia\\
	$^{6}$Instituto de Astronomia, Universidad Catolica del Norte, Av. Angamos 0610, Antofagasta 1270709, Chile\\
	$^{7}$Central (Pulkovo) Observatory, Pulkovskoe Shosse 65, Saint-Petersburg, Russia 196140
}

\date{Accepted XXX. Received YYY; in original form ZZZ}
\pubyear{2015}

\begin{document}
	
\pagerange{\pageref{firstpage}--\pageref{lastpage}} \pubyear{2018}
	
\maketitle
	
\label{firstpage}

\begin{abstract}
The determination of fundamental parameters of stars is one of the 
main tasks of astrophysics. For magnetic chemically peculiar stars, this problem is complicated 
by the anomalous chemical composition of their atmospheres, which requires 
special analysis methods. We present 
the results of the effective temperature, surface gravity, abundance and radius determinations for three CP stars 
HD~188041, HD~111133, and HD~204411. Our analysis is based on a self-consistent 
model fitting of high-resolution spectra and spectrophrotometric observations over a wide wavelength range, taking 
into account the anomalous chemical composition of atmospheres and the 
inhomogeneous vertical distribution for three chemical elements: Ca, Cr, and Fe.
For two stars, HD~188041 and HD~204411, we also performed interferometric
observations which provided us with the direct estimates of stellar radii.  
Comparison of the radii determined from the analysis of spectroscopic/spectrophotometric observations with direct 
measurements of the radii by interferometry methods for seven CP stars shows that the radii agree 
within the limits of measurement errors, which proves indirect spectroscopic analysis capable of proving
reliable determinations of the fundamental parameters of fainter Ap stars that are not possible to study
with modern interferometric facilities.
\end{abstract}


\section{Introduction}\label{intro}
Magnetic chemically peculiar (Ap) stars represent a group of stars of spectral types B5 to F5 with  global 
magnetic fields varying from few tens of Gauss 
\citep{2007A&A...475.1053A} to tens of kiloGauss \citep[see 
catalogue][]{2008AstBu..63..139R}. Magnetic fields have predominantly poloidal structure.   
While the Ap stars seem to have the same temperature, mass, 
luminosity and hydrogen line profiles as the normal Main Sequence (MS) stars, they show atmospheric abundance anomalies. 
As a rule, element abundance excess of up to 1-4 orders of magnitude compared to the solar abundances is observed for elements 
heavier than oxygen, while light elements, He, C, N, and O, are underabundant relative to the  
solar values \citep[see review][]{1991IAUS..145..149R}. 

\citet{1970ApJ...160..641M} proposed a mechanism of macroscopic diffusion for developing abundance anomalies in 
the atmospheres of CP stars.  In their atmospheres the diffusion of 
atoms and ions of a chemical element occurs under the combined action of gravitational settling and the radiation pressure that
act in opposite directions. If the 
gravitational pressure prevails, then the elements diffuse into the deep layers of the 
atmosphere of the star. In the opposite case, we have a directed flux of 
particles in the upper atmosphere. The drift of particles occurs with respect to the buffer gas, hydrogen. The element separation results in the creation of vertical abundance gradients that
produce the observed abundance anomalies. It changes the atmospheric structure via the line and continuum opacities, which leads to the change of the
observed energy distribution compared to the chemically normal stars. Therefore the standard photometric and spectroscopic calibrations developed 
for the determination of fundamental parameters in normal stars are often inapplicable in case of Ap stars.  
In Ap stars, the presence of significant individual anomalies of the 
chemical composition requires the detailed study of stellar chemistry to construct an adequate model atmosphere that can predict
the observed flux distribution. Self-consistent procedure of simultaneous spectrum and flux distribution modelling  
was first proposed by \citet{2009A&A...499..851K} for one of the brightest Ap stars $\alpha$~Cir. Then it was applied to a few other Ap stars
\citep{2009A&A...499..879S, 2010A&A...520A..88S, 2011MNRAS.417..444P, 2013A&A...551A..14S, 2013A&A...552A..28N}. The success of the proposed 
procedure was verified by the direct measurements of stellar radii in five Ap stars by means of interferometry \citep{2008MNRAS.386.2039B,2010A&A...512A..55B, 2011A&A...526A..89P, 2013A&A...559A..21P, 2016A&A...590A.117P}.

In this paper we extend the detailed self-consistent spectroscopic analysis to three other Ap stars, HD~188041 (V1291 Aquilae), HD~111133 (EP Virginis), 
and HD~204411. For two of them, HD~188041 and HD~204411, we performed interferometric observations and derived their radii.  
Spectroscopic observations and their analysis are presented in Section~\ref{spec}, 
and interferometric analysis is given in Section~\ref{interferometry}.


\section{Previous analysis}
\noindent
The investigated objects HD~188041 (V1291 Aquilae), HD~111133 (EP Virginis), and HD~204411 are well-studied
Ap stars of spectral class A1~SrCrEu, A6~SrCrEu, and A6~SrCrEu respectively \citep{2009A&A...498..961R}. 
The most anomalous element in their atmospheres is chromium and many of rare-earth elements (REE) which is typical
signature of Ap stars \citep{2004A&A...423..705R}.

Fundamental parameters (\teff, \lgg) of the investigated stars were derived in numerous studies in the past using different methods.
\citet{2008A&A...491..545N} (see the references therein) compiled effective temperature determinations for different groups of chemically peculiar stars. Evolutionary state of Ap stars was studied by \citet{2006A&A...450..763K} who derived effective temperatures and luminosities of 150 stars using the calibrations of Geneva photometry and improved Hipparcos parallaxes \citep{2007A&A...474..653V}. These data allow us to estimate stellar radii as well. 
Both investigations include our program stars.
HD~111133 and HD~188041 were analysed for Ca isotopic anomaly and stratification by \citet{2008A&A...480..811R}, where atmospheric parameters  \teff\, and \lgg\, were also determined. For HD~204411 fundamental parameters were derived from   
detailed spectroscopic and stratification study \citep{2005A&A...438..973R} using high-resolution SARG spectrum 
and Adelman's spectrophotometry \citep{1989A&AS...81..221A}. 
The authors used  Hipparcos parallax $\pi=8.37(53)$~mas \citep{1997ESASP1200.....E}. 
Position of HD~204411 on the H-R diagram shows that the star finished its main-sequence life. 
Although being detailed, the spectroscopic analysis of HD~204411 was not fully self-consistent.
Previous determinations of fundamental parameters of program stars are collected in Table~\ref{literat-param}.
Here and throughout the paper an error of the measurement in the last digits is given in parentheses. 

\begin{table*}
	\label{refs}
	\caption{Literature data for the fundamental parameters of program stars. }
	\small
	\centering
	\begin{tabular}{l c|c|c|c|l}
		\hline \hline HD  & \teff & \lgg & $\log(L/L_{\odot})$ & $R/R_{\odot}$ & Ref.\\
		\hline    & 9850(220) &          &          &        & {\citet{2008A&A...491..545N}}\\
		HD~111133 & 9930(250) &          & 1.92(13) & 3.09   & {\citet{2006A&A...450..763K}}\\
		          & 9930~~~~~~~~~& 3.65     &          &        & {\citet{2008A&A...480..811R}}\\
		\hline    & 8580(550) &          &          &        & {\citet{2008A&A...491..545N}}\\
		HD~188041 & 8430(200) &          & 1.55(07) & 2.80   & {\citet{2006A&A...450..763K}}\\
		          & 8800~~~~~~~~~& 4.00     &          &        & {\citet{2008A&A...480..811R}}\\
		\hline    & 8510(170) &          &          &        & {\citet{2008A&A...491..545N}}\\
		HD~204411 & 8750(300) &          & 1.95(06) & 4.12   & {\citet{2006A&A...450..763K}}\\
	              & 8400(200) & 3.50(10) & 2.01(10) & 4.6(2) & {\citet{2005A&A...438..973R}}\\
		\hline 
	\end{tabular}		 
\label{literat-param}
\end{table*}

In this work we extend the analysis of these stars by employing self-consistent abundance and stratification study
based on dedicated model atmospheres and available spectroscopic and (spectro-)photometric observations, as described
below.

\section{Spectroscopy}\label{spec}

\subsection{Observations}\label{obs}
\noindent
High-resolution spectra of HD~111133 and HD~188041 were obtained with the UV-Visual Echelle Spectrograph (UVES)
attached at the ESO VLT (program ID 68.D-0254). 
The resolving power of the spectrograph is R=$\lambda/\Delta\lambda$=80\,000, spectral region covered is  3100 to 10000~\AA. 
Details of the observations and data processing are given in \citet{2008A&A...480..811R}. 
HD~204411 was analysed using an echelle spectrum (R=164\,000, spectral region  4600--7900 \AA) obtained with the high resolution spectrograph ({\it SARG}) 
attached to the 3.55-m {\it Telescopio Nazionale Galileo} at the Observatorio del Roque de los Muchachos (La Palma, Spain). 
A detailed description of the data and data processing is given in \citet{2005A&A...438..973R}.
 
\subsection{Self-consistent spectroscopic analysis}\label{procedure}
\noindent
To accurately derive fundamental parameters -- effective temperature, surface gravity, radius, and luminosity -- of a magnetic CP star 
we applied an iterative method proposed by \citet{2009A&A...499..851K} for the roAp star $\alpha$~Cir and then used for few other 
Ap stars (see Section~\ref{intro}). 
Briefly, the basic steps of this method include the repeated stratification and abundance analysis used in the calculations of model atmosphere 
and theoretical SED which is compared to the observed one. SED fit also allows to estimate stellar radius provided that the stellar parallax is known. 
Iterations continue until fundamental parameters (\teff, \lgg, $R$, chemistry) used in SED fitting converge.
Thus, we get abundance pattern which is consistent with the physical
parameters of the final stellar atmosphere.
Model atmosphere calculations were done with the \textsc{LLmodels} code \citep{2004A&A...428..993S}.
As starting parameters in iteration procedure, we used model parameters from \citet{2008A&A...480..811R} 
for HD~111133, HD~118041 and from \citet{2005A&A...438..973R} for HD~204411. 
The details of our SED analysis will be given in Sect.~\ref{sec:sed-fit} and
below we only describe the steps of our self-consistent spectroscopic analysis.

\subsubsection{Abundance analysis}\label{abund}
\noindent
The abundances of chemical elements were evaluated using two different methods. Here and everywhere element abundance is given as 
$\log(N_{el}/N_{tot})$, where $N_{el}$ is a number of atoms of a considered element and $N_{tot}$ is a total number of atoms.  
A fast method of element abundance determination uses measured equivalent widths of the corresponding spectral lines.
For that we utilized the {\sc widthV} code \citep{2002A&A...384..545R}. However, it is difficult to apply simple equivalent widths calculations to magnetic stars 
due to magnetic splitting and intensification of spectral lines. Therefore, a special version of the {\sc widthV} code called {\sc WidSyn} 
was developed \citep{2013A&A...551A..14S}. The program provides an interface to magnetic spectrum synthesis code {\sc SYNMAST} \citep[for a more detailed 
description see][]{2007pms..conf..109K, 2010A&A...524A...5K}
where theoretical equivalent widths are calculated from the full polarized radiative transfer spectrum synthesis. 
We assumed surface magnetic field to be constant with depth and defined by radial  and meridian components of magnetic field vector whose 
modulus \bs\ is derived from the magnetic splitting of spectral lines in nonpolarized observed spectrum. 
The observed \bs\ values are taken from \citet{2008A&A...480..811R} for HD~111133 and HD~188041, from \citet{2005A&A...438..973R} for HD~204411 
and are given in Table~\ref{HRD-table}. Line atomic parameters 
were extracted from the Vienna Atomic Line Database {\sc VALD} 
\citep{1999A&AS..138..119K} in its 3d release {\sc VALD3}  \citep{2015PhyS...90e4005R}.

\begin{figure}
	\centering
	\includegraphics[width=0.45\textwidth,clip]{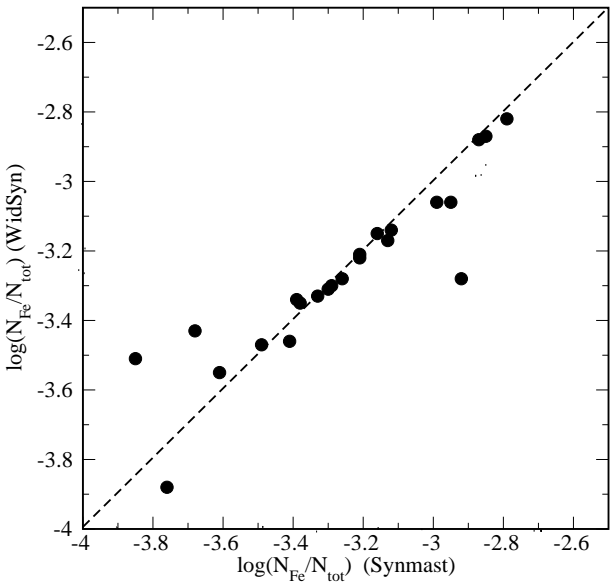}
	\caption{Comparison between Fe abundances derived from the equivalent widths ({\sc WidSyn} code) and by magnetic spectrum fitting 
    ({\sc SYNMAST} code).}
	\label{EqW}
\end{figure}

Another, more accurate method to measure element abundance is based on a direct fitting of the theoretical 
profiles of individual spectral lines by varying parameters  
\vs, \vr\ (stellar projected rotation and radial velocity, respectively), two components of the magnetic field vector,  and element abundance. 
It was done with {\sc SYNMAST} code implemented in the visualization program \bmag\
\citep{2018ascl.soft05015K}. The direct spectrum synthesis takes into account 
possible blends, since the lines are often blended in spectra of CP stars, and the 
abundance estimation by equivalent widths may not be very accurate. We performed a comparative analysis of the Fe (HD~111133) and Nd (HD~188041) 
abundances derived by both methods using 23 Fe\ione,\ii\ lines and 11 Nd\iii\ lines. Figure~\ref{EqW} shows this comparison for Fe lines. Due to the careful choice of spectral lines the abundances derived by both methods agree rather well.
The corresponding Fe abundance is -3.26(29) ({\sc WidSyn}) versus -3.26(26) ({\sc SYNMAST}). A similar satisfactory agreement was obtained for Nd abundance.

Based on the results of our investigation we prefer to use equivalent width method for abundance analysis in most cases.
The mean atmospheric abundances of 36 elements were derived at each iteration. For 17 elements this was done using
spectral lines of two ionization stages.
The isotopic structure of Ba~II and Eu~II was taken into account, however we neglected the hyperfine splitting (hfs) because of the complex 
interaction between the hyperfine and magnetic splittings \citep{1975A&A....45..269L}. It means that abundances of a few elements with the odd isotopes 
(Mn, Co, Pr, Eu, Tb) may be overestimated. The only exception is HD~204411 where we neglected the magnetic field in abundance analysis. 
For this star we employed profile fitting method when necessary and used hfs data from \citet{2005ApJS..157..402B, HSR} 
(Mn\ione, \ii), \citet{P} (Co\ione), \citet{VAHW} (Ba\ii).

Average abundances for the final atmospheric models are given 
in Table~\ref{MeanAbunds}. Model parameters are presented in Section~\ref{flux}. 
The last column of the table presents the current solar photospheric abundances \citep{2009ARA&A..47..481A, 
	2015A&A...573A..25S, 2015A&A...573A..26S,
	2015A&A...573A..27G}.

 \begin{table}
	\caption{The mean values of chemical abundance \abun, calculated from equivalents widths of N lines for model atmospheres 8770g42 (HD~188041),  
		9800g40 (HD~1111333), and 8300g36 (HD~204411).
		Values derived from one line are marked with a colon. The standard deviation is given in parentheses. 
		Solar photospheric abundances are given in the last column.}
	\scriptsize
	\centering
	\begin{tabular}{l c|r|c|r|c|r|c}
		\hline \hline Ion & \multicolumn{4}{c|}{Abundance \abun} \\ 
		\hline & HD~188041 & N & HD~111133 &  N &  HD~204411 & N &   Sun\\
		\hline 
		C\ione & -4.17(29) & 3 & -4.25(54) & 3 &~-4.48(15) & 5&~-3.61 \\ 
		N\ione & -4.94(25) & 2 & -5.10:~~~ & 1 &~-4.15(05) & 2&~-4.21 \\ 
		O\ione & -4.60(36) & 4 & -3.93(43) & 7 &~-3.71(07) & 4&~-3.35 \\ 
		Na\ione& -5.62(15) & 2 & -5.68(70) & 2 &~-5.63(07) & 2&~-5.83 \\ 
		Mg\ione& -4.34(80) & 2 & -4.42(18) & 4 &~-4.60(20) & 6&~-4.45 \\ 
		Mg\ii  & -4.28(13) & 2 & -4.49(52) & 5 &~-4.62(49) & 3&~-4.45 \\ 
		Al\ione& -4.50:~~~ & 1 &           &   &	          &  &~-5.61 \\
		Al\ii  & -4.57(13) & 2 & -5.90(01) & 2 &	          &  &~-5.61 \\ 
		Si\ione& -3.90(19) & 5 & -4.09(60) & 9 &~-4.33(15) & 6&~-4.53 \\ 
		Si\ii  &           &   & -4.65(35) & 4 &~-4.12(03) & 2&~-4.53 \\ 
		S\ione & -3.44:~~~ & 1 &           &   &~-5.12(23) & 2&~-4.92 \\ 
		S\ii   &           &   & -5.09(50) & 3 &	          &  &~-4.92 \\
		Ca\ione& -4.69(58) &11 & -5.98:~~~ & 1 &~-5.54(05) & 5&~-5.72 \\ 
		Ca\ii  & -4.96(32) & 2 & -6.87(20) & 3 &~-5.05(11) & 5&~-5.72 \\ 
		Sc\ii  & -8.32(70) & 5 &           &   &~-9.68:~~~ & 1&~-8.88 \\ 
		Ti\ione&           &   &           &   &~-6.81(11) &13&~-7.11 \\ 
		Ti\ii  & -6.45(54) &14 & -6.51(42) &29 &~-6.63(11) &15&~-7.11 \\ 
		V\ione & -5.60(23) & 2 &           &   &	          &  &~-8.15 \\
		V\ii   & -6.80(47) & 6 & -8.05(19) & 9 &~-8.87(09) & 2&~-8.15 \\ 
		Cr\ione& -3.97(57) &10 & -3.71(31) &66 &~-5.18(16) &37&~-6.42 \\ 
		Cr\ii  & -3.81(54) &15 & -3.88(43) &248&~-4.90(19) &36&~-6.42 \\ 
		Mn\ione& -4.94(62) & 9 & -4.96(28) & 9 &~-6.25(13) & 9&~-6.62 \\ 
		Mn\ii  & -4.99(88) & 5 & -4.97(39) &51 &~-5.91(18) &14&~-6.62 \\ 
		Fe\ione& -3.60(35) &19 & -3.24(27) &153&~-4.10(20) &79&~-4.57 \\ 
		Fe\ii  & -3.21(40) &52 & -3.24(32) &282&~-3.64(34) &69&~-4.57 \\ 
		Co\ione& -5.05(63) & 9 & -5.09(47) &14 &~-6.65(20) & 3&~-7.11 \\ 
		Co\ii  & -5.50:~~~ & 1 & -5.48(52) &12 &           &  &       \\
		Ni\ione& -5.81(63) & 5 & -5.77(76) &10 &~-5.96(14) &27&~-5.84 \\ 
		Ni\ii  &           &   & -6.18(57) & 3 &~-5.42:~~~ & 1&~-5.84 \\ 
		Cu\ione& -6.88:~~~ & 1 &           &   &~-8.26:~~~ & 1&~-7.86 \\ 
		Sr\ione& -5.81:~~~ & 1 &           &   &~-8.99:~~~ & 1&~-9.21 \\ 
		Sr\ii  & -5.97(12) & 2 & -7.07(51) & 3 &	          &  &~-9.21 \\ 
		Y\ii   & -9.23(40) & 4 & -8.83(22) & 3 &-10.24(14) & 4&~-9.83 \\ 
		Zr\ii  & -8.25(57) &10 & -8.57(77) &10 &~-9.12(28) & 2&~-9.45 \\ 
		Nb\ii  & -8.09(73) & 2 &           &   &	          &  &-10.57 \\ 
		Mo\ii  & -7.72(17) & 2 &           &   &	          &  &-10.16 \\
		Ba\ii  & -9.12(57) & 4 & -8.93(39) & 3 &~-9.45(23) & 3&~-9.79 \\ 
		La\ii  & -8.29(23) & 6 & -7.97(60) &18 &	          &  &-10.93 \\
		Ce\ii  & -7.62(31) &14 & -7.46(62) &27 &-10.46:~~~ & 1&-10.46 \\ 
		Ce\iii & -6.08(35) & 6 & -7.14(24) & 4 &	          &  &-10.46 \\
		Pr\ii  & -7.91(38) & 4 & -7.90(41) & 6 &	          &  &-11.32 \\ 
		Pr\iii & -8.47(30) & 9 & -8.30(43) &25 &	          &  &-11.32 \\ 
		Nd\ii  & -8.33(56) & 3 & -7.64(40) &24 &~-9.80(22) & 3&-10.62 \\ 
		Nd\iii & -7.87(28) &12 & -8.32(44) &33 &-10.14:~~~ & 1&-10.62 \\ 
		Sm \ii & -8.52(30) & 6 & -8.06(35) &15 &	          &  &-11.09 \\
		Sm\iii & -7.07:~~~ & 1 & -8.46(04) & 2 &	          &  &-11.09 \\
		Eu\ii  & -7.41(24) & 5 & -8.93(58) & 6 &-11.19:~~~ & 1&-11.52 \\ 
		Eu\iii & -5.29(53) & 4 & -6.68(29) & 7 &	          &  &-11.52 \\
		Gd\ii  & -7.48(71) & 8 & -7.38(46) &25 &	          &  &-10.96 \\
		Tb\ii  & -7.91:~~~ & 1 &           &   &	          &  &-11.70 \\
		Tb\iii & -8.85(36) & 4 & -8.44(50) & 1 &	          &  &-11.70 \\
		Dy\ii  & -8.95(48) & 4 & -7.88(45) &13 &	          &  &-10.94 \\
		Er\ii  & -9.38(45) & 4 & -8.20(45) & 7 &	          &  &-11.11 \\
		Er\iii & -8.06:~~~ & 1 & -8.56(56) & 5 &	          &  &-11.11 \\
		Tm\ii  & -8.83(1.42)& 3 & -9.03:~~~ & 1 &          &  &-11.93 \\
		Yb\ii  & -8.07(44) & 5 & -8.09(37) &10 &	          &  &-11.19 \\			   
		\hline 
	\end{tabular}		 
	\label{MeanAbunds}
\end{table}

\begin{figure*}
	\centering
	\includegraphics[width=0.8\textwidth]{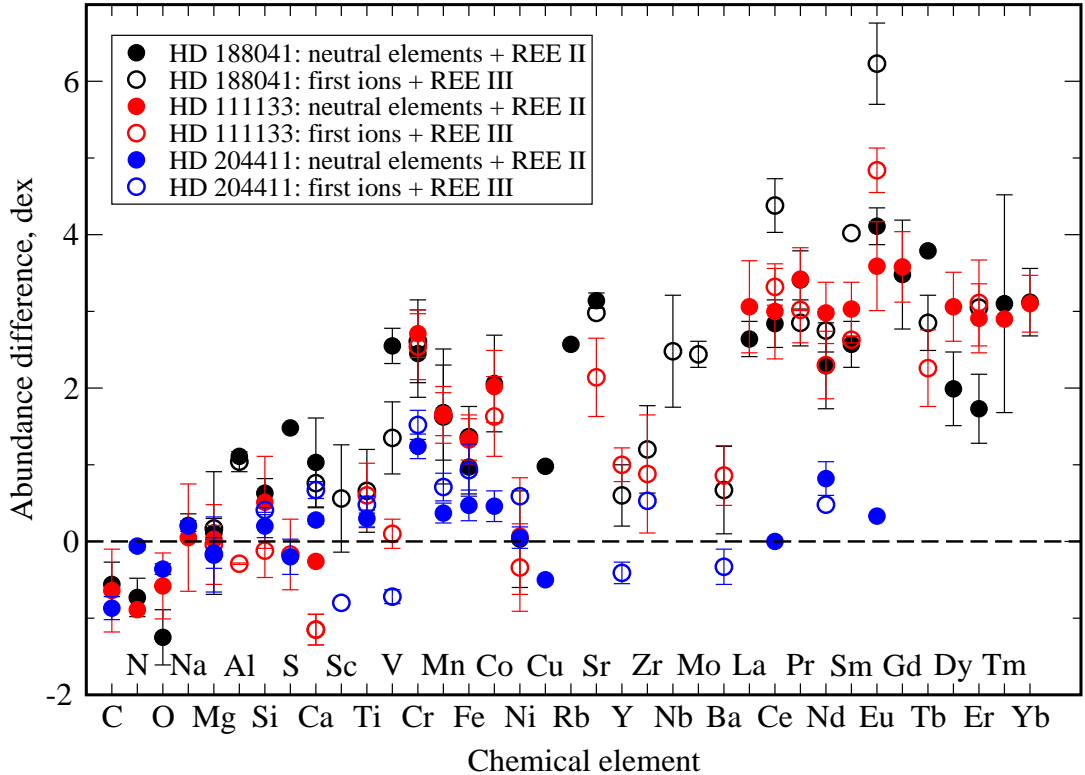}
	\caption{Element abundances relative to solar values in the atmospheres of
		HD~188041 (black symbols), HD~111133 (red symbols), and HD~204411 (blue symbols). 
        Abundances derived from the lines of the neutral species (C to Ba) and of the first ions (La to Yb) are shown with filled circles, and
        those derived from the lines of the first (C to Ba) and the second (La to Yb) ions are shown with open circles.}
	\label{abund-vs-sun}
\end{figure*}

Figure~\ref{abund-vs-sun} shows the mean element abundances in the 
atmospheres of investigated stars relative to the solar photospheric values. Abundances derived from the lines of
consecutive ionization stages (neutral and first ions or first and second ions for REE) are shown by filled and open circles, respectively. The abundances in
HD~111133 and HD~188041 are similar within the errors of the 
determination, and the abundance patterns correspond to the general observed anomalies in Ap-stars: 
CNO deficiency, practically solar Na and Mg abundance, 1-2~dex overabundance of the iron peak elements, and a large 
excess of the rare-earth elements (REEs). However, the abundance pattern in HD~204411 is different. CNO deficiency is the same, iron peak elements 
are less overabundant, and the REEs are close to the solar values. We discuss this issue in Section~\ref{evol}.

Inspecting Table~\ref{MeanAbunds} one notices rather large abundance difference derived from the lines of consecutive ionization stages of some elements, e.g. Ca, Cr, Fe, Eu. Usually it may be an indication of abundance stratification in stellar atmosphere \citep{2003IAUS..210..301R}. 
There is no significant violation of the ionization equilibrium of 
those rare-earth elements (Ce, Pr, Nd, Sm) for which the abundance is determined 
more or less reliably by several lines of different ionization stages. The 
exception is Eu, where the Eu\iii\ lines provide more than an order of 
magnitude higher abundance compared to the Eu\ii\ lines. This behaviour 
is typical for most Ap stars in the effective temperature range of 7000--10000~K 
\citep{2017AstL...43..252R}. 

As was mentioned above, some elements may have inhomogeneous vertical distribution, which changes atmospheric structure through the variable line opacity.
We performed stratification analysis of iron, chromium and calcium, because the lines of these elements (in particular, Fe and Cr) dominate 
the observed spectrum.

\subsubsection{Stratification study}\label{strat}
\noindent

Element  stratification  was calculated using the {\sc DDaFit} IDL-based automatic procedure \citep{2007pms..conf..109K}.
Theoretical stratification calculations show that, as a first approximation, the stratification profile of chemical elements
can be represented by a step 
function \citep{1992A&A...258..449B}. This function is described by four parameters: element abundances in the 
upper and lower atmospheric layers, the position of the abundance jump and the width of this jump. These parameters are optimized
to provide the best fit to the observed profiles of spectral lines formed at different atmospheric layers.  Spectral synthesis is carried out 
with the {\sc SYNMAST} program, and the possible contribution of 
neighboring lines is taken into account. The success of the stratification study strongly depends on the choice of spectral lines. 
The chosen lines should be sensitive to possible abundance variations at significant part of atmospheric layers. 
The list of the lines is given in Table~\ref{list-strat} and is available in electronic version of the paper. Here we present
only a part of this table.

\begin{table*}
	\centering
	\small
	\caption{A list of spectral lines used for the stratification 
		calculations for stars HD~188041, HD~111133, and HD~204411 with the excitation
		potential (eV), oscillator strength ($\log\,{gf}$) and Stark damping parameter
		($\log\,\gamma_{\rm St}$) listed. }
	\begin{tabular}{l|c c c c|c|c|c|}
		\hline Ion & $\lambda$ (\AA) & $E_{\textnormal i}$ (eV) &$\log\,{gf}$ & $\log \gamma_{\rm St}$ & 
		HD~188041 & HD~111133& HD~204411\\ \hline
	Fe II &  4504.343 & 6.219 & -3.250 &  -6.530 &$\surd$&       &       \\
    Fe II &  4508.280 & 2.855 & -2.250 &  -6.530 &       &$\surd$&       \\
    Fe II &  4610.589 & 5.571 & -3.540 &  -6.530 &$\surd$&$\surd$&       \\
    Fe II &  4631.867 & 7.869 & -1.945 &  -5.830 &$\surd$&$\surd$&       \\
    Fe II &  4975.251 & 9.100 & -1.763 &  -6.530 &       &$\surd$&       \\
    Fe II &  4976.006 & 9.100 & -1.599 &  -6.530 &$\surd$&       &       \\
    Fe I  &  5014.942 & 3.943 & -0.303 &  -5.450 &$\surd$&$\surd$&       \\
    Fe II &  5018.435 & 2.891 & -1.220 &  -6.530 &$\surd$&$\surd$&$\surd$\\
    Fe II &  5018.669 & 6.138 & -4.010 &  -6.537 &       &       &$\surd$\\
    Fe I  &  5019.168 & 4.580 & -2.080 &  -5.970 &       &       &$\surd$\\
    ......& ......    & ......& ...... & ......  & ..... & ......& ......\\
 \hline 
            \multicolumn{8}{p{.6\textwidth}}{This table is available in its entirety in a machine-readable form in the online journal. A portion is shown here for guidance regarding its form and content. Stark damping parameters are given for one perturber (electron) for T=10000~K.
} \\
\end{tabular}
\label{list-strat}
\end{table*}

\begin{figure*}
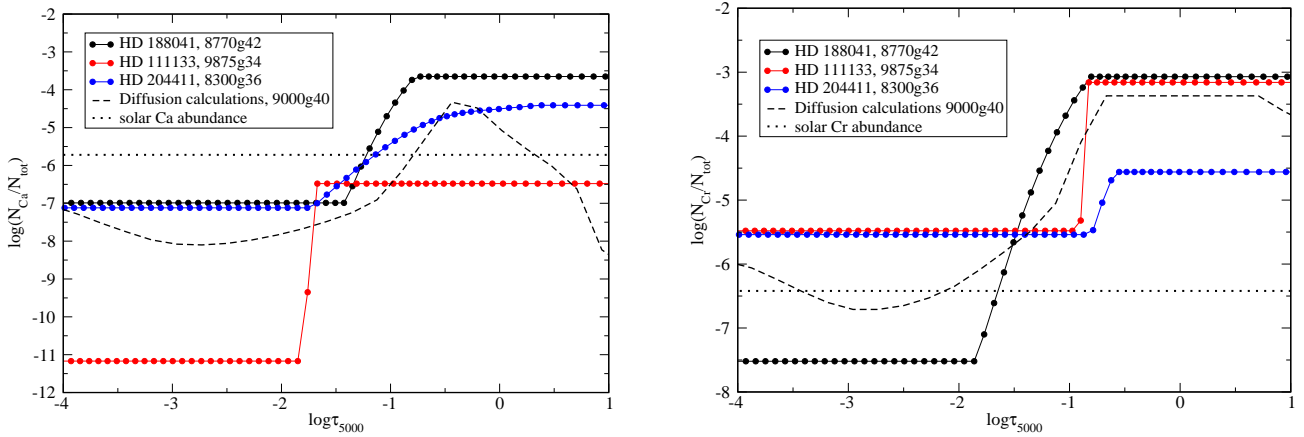

	\includegraphics[width=0.45\textwidth]{Ca_distr_cor.eps}\hspace{10mm}\includegraphics[width=0.45\textwidth]{Cr_distr_cor.eps}
	\caption{Distribution of Ca (left panel) and Cr (right panel) in the atmospheres of HD~188041 (black filled circles), HD~111133 (red filled circles), and
	HD~204411 (blue filled circles). Theoretical predictions for the model \teff=9000~K, \lgg=4.0 are shown by dashed line. 
    Solar abundances are indicated by dotted line.}
	\label{Ca-Cr}
\end{figure*}

\begin{figure}
	\includegraphics[width=0.45\textwidth]{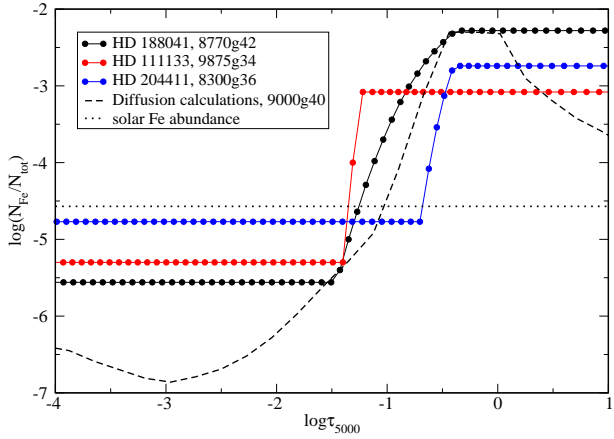}
	\caption{The same as in Fig.~\ref{Ca-Cr}, but for Fe.}
	\label{Fe}
\end{figure}

Stratification calculations began with the homogeneous abundance distribution corresponding to the average atmospheric abundances derived 
as described in Section~\ref{abund}. 
{\sc DDaFit} procedure is repeated for each iteration, every time 
using model atmosphere with improved atmospheric parameters \teff, \lgg, and abundance patterns which are refined through 
the fitting of spectral energy distribution (see Section \ref{flux}).  
In Fig.~\ref{Ca-Cr} and Fig.~\ref{Fe}  we show final Ca, Cr, and Fe stratifications.

All the three elements have abundance jumps in the atmospheres of investigated stars, and these jumps are located at the optical depths 
close to the predictions of the diffusion theory. It is demonstrated by the comparison of the empirical stratifications with the theoretical 
diffusion calculations for model \teff=9000~K, \lgg=4.0  \citep{2005JRASC..99T.139L, 2009A&A...495..937L}. 
However, in the atmosphere of the most evolved star HD~204411 abundance jumps seem to be smaller, in particular, for Cr. 
All the three stars show large Fe and Cr overabundance in the deep atmospheric layers where high-excitation lines are formed. 
The most unusual result is Ca deficiency throughout the whole atmosphere of HD~111133, which is not predicted by the diffusion theory. 

The results of our analysis do not differ  much from the derived Ca stratification in HD~188041 and HD~111133 \citep{2008A&A...480..811R}, 
and from Ca-Cr-Fe stratifications derived by \citet{2005A&A...438..973R}. It is not surprising because the cited studies 
were based on the same observed spectra. 
Stratification analysis allows us to derive more accurately stellar rotational velocities.
The best fits of the calculated line profiles to the theoretical ones were obtained for  \vs=2~\kms\ in
HD~188041 and for \vs=6~\kms\ in HD~111133.

\section{Analysis of spectral energy distribution}\label{sec:sed-fit}

\subsection{Observations}\label{sec:sed-obs}
\noindent
Spectral energy distribution (SED) in absolute units for all the three stars was constructed from UV wide-band fluxes obtained with TD1 space mission
and extracted from TD1 catalogue \citep{1978csuf.book.....T}, from optical spectrophotometry \citep{1976ApJS...32....7B,1989A&AS...81..221A}, 
and from the near infrared 2MASS (2Micron All-Sky Survey) catalogue  which contains an overview of the entire sky at $J$-band (1.25 $\mu m$), $H$-band (1.65 $\mu m$), 
and $K$-band (2.17 $\mu m$).
The spectrophotometric magnitudes by \citep{1989A&AS...81..221A} were converted to fluxes using Vega calibrations
given in \citet{1975ApJ...197..593H}.
We convert 2MASS magnitudes to absolute fluxes using filter values and zero points defined in \citet{2003AJ....126.1090C}.

For HD~188041, we additionally utilized fluxes between $1900$~\AA\ and $3000$~\AA\ observed 
with International Ultraviolet Explorer (IUE)\footnote{\tt http://archive.stsci.edu/iue/}.

Along with the low-resolution spectrophotometric data by \citep{1989A&AS...81..221A}, 
for HD~204411 we made use of the medium resolution spectroscopic observations with
the Boller \& Chivens long-slit spectrograph mounted at $2.1$\,m telescope at the Observatorio
Astron\'omico Nacional at San Pedro M\'artir, Baja California,
M\'exico (OAN SPM)\footnote{\tt http://www.astrossp.unam.mx/oanspm/}. 
The data for HD~204411 and several other Ap stars was obtained during July, 2011.
In order to make precise spectrophotometry, the slit was opened as wide as possible to cover
the star's photometric profile together with
its far wings and surrounding sky background. With this configuration
spectral resolution of about $8$~\AA\ and
simultaneously registered wavelength range from $3700$~\AA\ to $7200$~\AA\ were achieved.

The wavelength calibration with an arc lamp spectrum obtained several
times per the observing night, spectrograph
flexion correction, and flat-fielding with a halogen lamp were made
in a standard manner. The spectra of the star were
flux-calibrated using spectrophotometric standard stars observed at
different zenith distances during the same night. In particular case
of HD~204411 we used two calibrator stars HD~192281 and BD+33d2642, respectively.

\subsection{The determination of fundamental stellar parameters.}\label{flux}
\noindent
To compare the observed and theoretical fluxes at different wavelengths (spectral energy distribution -- SED) 
one needs to calculate irradiation from the surface of a star located at some distance.
This irradiation depends on stellar atmospheric parameters, stellar radius, and the distance to the star. 
The distance is defined by stellar parallax. We compute theoretical fluxes at different wavelengths from the adopted 
model atmosphere. We optimize stellar radius, \teff, and \lgg\ for a given abundance pattern and stratification to reach the best fit 
to the observed energy distribution. 
This approach allows us to refine the atmospheric parameters and the radius of the star simultaneously.

Because we utilize datasets that come from different space and ground-based missions, the number of observed points that we fit can differ
from only a few  (e.g., 2MASS) to hundreds (e.g., IUE) depending on wavelength domain. Therefore, in order to ensure that each wavelength domain
equally contribute to the final fit, we weight data in UV, Visual, and IR spectral ranges
by corresponding number of observed points. We chose three weighting regions: $0$-$4200$\AA\AA, $4200$-$11000$\AA\AA, $11000$-$\infty$\AA\AA.
The first region is useful for the estimation of stellar gravity because it includes the Balmer jump whose amplitude
is sensitive to the atmospheric pressure for stars of early spectral types. The second and the third regions are mostly sensitive
to the stellar effective temperature. Altogether, having observations in UV, Visual, and IR helps to constrain accurate atmospheric
parameters \citep{2013A&A...551A..14S}.

It is known that CP stars exhibit variability of their fluxes with the rotation period as observed in different
photometric filters \citep[e.g.,][]{1973BAAS....5..325M,1989A&AS...81..221A}.
This variability is caused by non-homogeneous distribution of chemical elements across the stellar surface (abundance spots)
which produces stellar brightness variation due to the changes in local atmospheric opacity, as explained in numerous
investigations \citep[see, e.g.,][]{2009A&A...499..567K,2010A&A...524A..66S,2012A&A...537A..14K,2015A&A...576A..82K}.
Among different observation data sets available to us only \citet{1989A&AS...81..221A} set
contains several scans per star taken at different rotation phases. However, the flux variability
in our sample stars reaches maximum of about $12$\%\ in HD~204411, 
$9$\%\ in HD~188041, and $6$\%\ in HD~111133, respectively.
We therefore conclude that abundance spots do not introduce strong biases in our parameters
estimation. Note that the amplitude of flux variation is the largest at short
wavelengths where the atmospheric opacity is the strongest. Unfortunately, the available UV fluxes
provided by the TD1 sattelite do not contain time resolved information.
Because in our SED fit we combine data obtained at differench epochs and with different missions and/or instruments
we could not study the impact of abundance spots on determination of atmospheric parameters in full detail.
Thus, in our analsys we minimize the impact of phase variability of stellar fluxes by averaging 
individual scans taken by \citet{1989A&AS...81..221A} for each star.

The parallax values were taken from the second GAIA release  DR2 \citep{2018A&A...616A...1G}. Regarding the total uncertainty $\sigma_{\rm ext}$ on each parallax, we have computed it using the formula provided by Lindegren et al. at the IAU colloquium\footnote{https://www.cosmos.esa.int/web/gaia/dr2-known-issues\#AstrometryConsiderations}:
\begin{equation}
    \sigma_{\rm ext} = \sqrt{k^2 \sigma_i^2 + \sigma_s^2}
\end{equation}
with $\sigma_i$ provided in the GAIA DR2 catalogue, $k$~=~1.08 and $\sigma_s$~=~0.021~mas for $G$~<~13.
We thus considered the values of 11.72(11)~mas (HD~188041), 5.21(07)~mas (HD~111133), and 8.33(12)~mas (HD~204411), respectively.
The adopted parallaxes differ from the original  \citep{1997ESASP1200.....E} and revised  \citep{2007A&A...474..653V} Hipparcos ones by 
5-6~\% for HD~188041 and HD~204411, and by more than 30~\% for HD~111133. 
For example, for HD~188041 the available parallax values are 11.79(93) \citep{1997ESASP1200.....E}, 
12.48(36) \citep{2007A&A...474..653V}, 11.95(97) \citep{2016A&A...595A...1G}, 
and 11.72(10) \citep{2018A&A...616A...1G}, 
while the corresponding values for HD~111133 are 6.23(93), 3.76(40), 5.95(73), and 5.21(06), respectively. 
Such a large scatter in parallax values for HD~111133 influences the radius estimate and, hence, the luminosity (see Sections~\ref{HD111133} and \ref{evol}).

Our three stars are located at the short distance from the Sun, therefore the corrections 
for interstellar reddening $A_v$ are small and were taken to be zero for HD~188041, 0.$^m$03 for HD~111133, and 0.$^m$016 for HD~204441. 
We calculated these values based on the reddening maps from \citet{2005AJ....130..659A}. 
It corresponds to the E(B-V) values of 0.$^m$009 (HD~111133) and 0.$^m$005 (HD~204411) if one takes a typical parameter $R_v$=3.1 in standard extinction low. 
In the last two years new 3D reddening maps were published based on the data from GAIA mission. 
Table~\ref{redden} collects E(B-V) data from four different sources: \citet{2005AJ....130..659A},  \citet{2018AA...616A.132L}\footnote{https://stilism.obspm.fr/}, 
\citet{2018MNRAS.478..651G}, and \citet{2018yCat.2354....0G}. One can easily see rather large dispersion between different sets. 
Colour excess E(B-V) extracted from the catalogs of \citet{2018yCat.2354....0G} provide the largest estimates, while the maps 
of interstellar dust in the Local Arm constructed by \cite{2018AA...616A.132L} using Gaia, 2MASS, and APOGEE-DR14 data give us 
the E(B-V) values not very different from Amores \& Lepine. \citet{2018MNRAS.478..651G} provide a code for E(B-V) calculations in different modes. 
We chose data calculated in mode 'median'. In order to account for interstellar reddening we used the \textsc{fm\_unred} routine from the IDL 
Astrolib package\footnote{https://idlastro.gsfc.nasa.gov/}
that uses reddening parameterization after \citet{1999PASP..111...63F}.

\begin{table}
	\centering
	\small
	\caption{Reddening data for HD~111133, HD~188041 and HD~204411.}
	\begin{tabular}{l c c c c c}
		\hline
		HD &  d, pc &\multicolumn{4}{c}{E(B--V)} \\
		   &        & 1 & 2 & 3 & 4\\		
		\hline
		HD~111133 & 191.9(2.1) & 0.010 & 0.017(020) & 0.027& 0.06(05)\\
		HD~188041 & ~85.3(0.7) & 0.000 & 0.007(017) & 0.012& 0.10(05)\\
		HD~204411 & 120.0(1.6) & 0.005 & 0.011(015) & 0.028& 0.07(05)\\
		\hline 
		\multicolumn{6}{p{.45\textwidth}}{1 -- \citet{2005AJ....130..659A}, 2 -- \citet{2018AA...616A.132L}, 3 -- \citet{2018MNRAS.478..651G}, 4 -- \citet{2018yCat.2354....0G}
		}\\		
	\end{tabular}
	\label{redden}	
\end{table}

\begin{figure}
	\includegraphics[width=0.45\textwidth,clip]{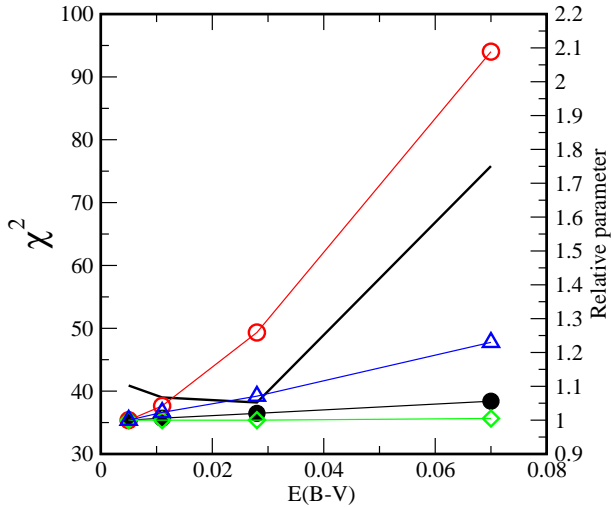}
	\caption{Fundamental stellar parameters of HD~204411 as a function of E(B-V). Black solid line shows $\chi^2$ of the SED fit (left y-axis), while variation of the relative values for \teff\, (black filles circles), \lgg\, (open red circles), radius (open green diamonds) and luminosity (open blue triangles) are marked on the right y-axis. The reference parameters derived with E(B-V)=0.$^m$005 are given in Table~\ref{HRD-table}.}
	\label{HD204411-red}
\end{figure}

We checked SED fitting for all E(B-V) from Table~\ref{redden}. Figure~\ref{HD204411-red} demonstrates 
the influence of the reddening on the derived stellar parameters \teff, \lgg, radius and luminosity in HD~204411. 
The similar picture was obtained for HD~188041. It is evident that our solution for two stars, HD~188041 and HD~204411, 
with the reddening data from \citet{2005AJ....130..659A}  provides the proper atmospheric parameters, 
and a slight increase in E(B-V) (first three points in Fig.~\ref{HD204411-red}) produces parameter variations within the uncertainties indicated in our work. The largest variations are observed for gravity, but in logarithmic scale it corresponds to the uncertainty 0.1~dex given in Table~\ref{HRD-table}.   

However, for HD~111133 the situation is opposite. The best fit to the SED was obtained for the largest reddening,
but the observed high resolution spectrum cannot be fit with the atmospheric parameters derived with this reddening values (see Section~\ref{HD111133}).

For all stars, models with a solar helium abundance and with the reduced by several orders helium abundance were constructed because
the helium deficiency is typical for peculiar stars. Our analysis showed that models with  normal helium abundance 
provide slightly better fit to the observed SED.
Final atmospheric parameters and stellar radii are obtained during the iterative process as described in Section~\ref{procedure}. 
Five iterations were required to reach the convergence for HD~188041, 3 iterations for HD~111133, and 2 iterations for HD~204411. 

\subsubsection{HD~188041}
\begin{figure*}
	\includegraphics[width=0.9\textwidth]{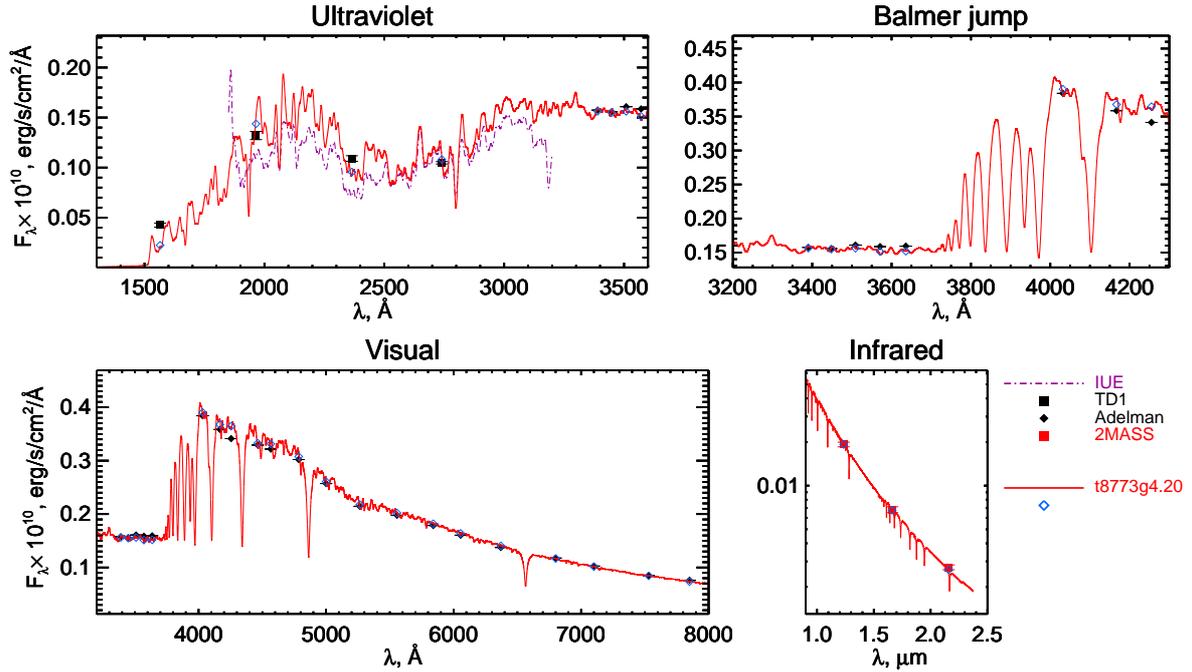}
	\caption{Comparison of the observed energy distribution (black filled squares and diamonds) with the theoretical flux 
	(red solid line) calculated for \textsc{LLmodels} atmosphere of HD~188041 with the parameters \teff=8770~K, \lgg=4.2. 
	Open blue diamonds show the theoretical fluxes convolved with the corresponding filters. 
    2MASS photometric colours are presented by filled red squares.}
	\label{sed-188041}
\end{figure*}

\noindent
Figure ~\ref{sed-188041} shows a comparison between the observed SED and the theoretical calculations
for the atmosphere of HD~188041. Using TD1 fluxes in UV wavelength domain we obtained the following best fit parameters:
\teff=8770(30)~K, \lgg=4.2, solar helium abundance, and radius $R/R_\odot$ = 2.39(2). However, we could not obtain good constraints for the
\lgg\, because our fitting algorithm was always favoring the largest value available in our model grid. On the contrary, when using IUE fluxes
we obtained \teff=8613(11)~K, \lgg=4.02(5)~K, $R/R_\odot$ = 2.44(1) and thus better estimates for the \lgg\, which agrees 
with what is expected for A-type main-sequence stars. The lower \teff\ obtained from fitting IUE flux is because the latter appears
to be systematically lower compared to TD1 flux.

\subsubsection{HD~111133}\label{HD111133}
\begin{figure}
	\includegraphics[width=0.45\textwidth]{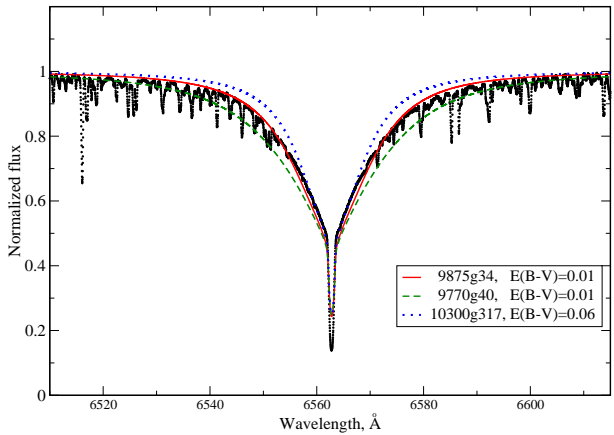}
	\caption{Comparison of the observed H$\alpha$ line profile in HD~111133 (black filled circles) with the synthetic profiles calculated using model atmospheres 9875g34 (red solid line), 9770g40 (dashed green line), and 10300g317 (blue dotted line).}
	\label{H-lines}
\end{figure}

\noindent
In all attempts to fit the observed SED with E(B-V)$\leq$0.$^m$03 we could not constrain
stellar \lgg. Using reddening from \citet{2018yCat.2354....0G} we got the best $\chi^2$ fit, however, model atmosphere with 
\teff=10300~K, \lgg=3.17 cannot reproduce hydrogen line profiles and Fe\ione/Fe\ii\, lines with any stratification. 
A detailed analysis carried out for HD~111133 with E(B-V)=0.$^m$010 --0.$^m$027 showed slight but not statistically significant
improvement of $\chi^2$ SED fit with the increase of E(B-V) for two fixed \lgg: 4.0 and 3.4. SED fit with fixed \lgg=4.0 provides slightly better
$\chi^2$, however \lgg=3.4 is preferred because it fits the  hydrogen lines better (see Fig.\ref{H-lines}).   

\begin{figure*}
	\includegraphics[width=0.9\textwidth]{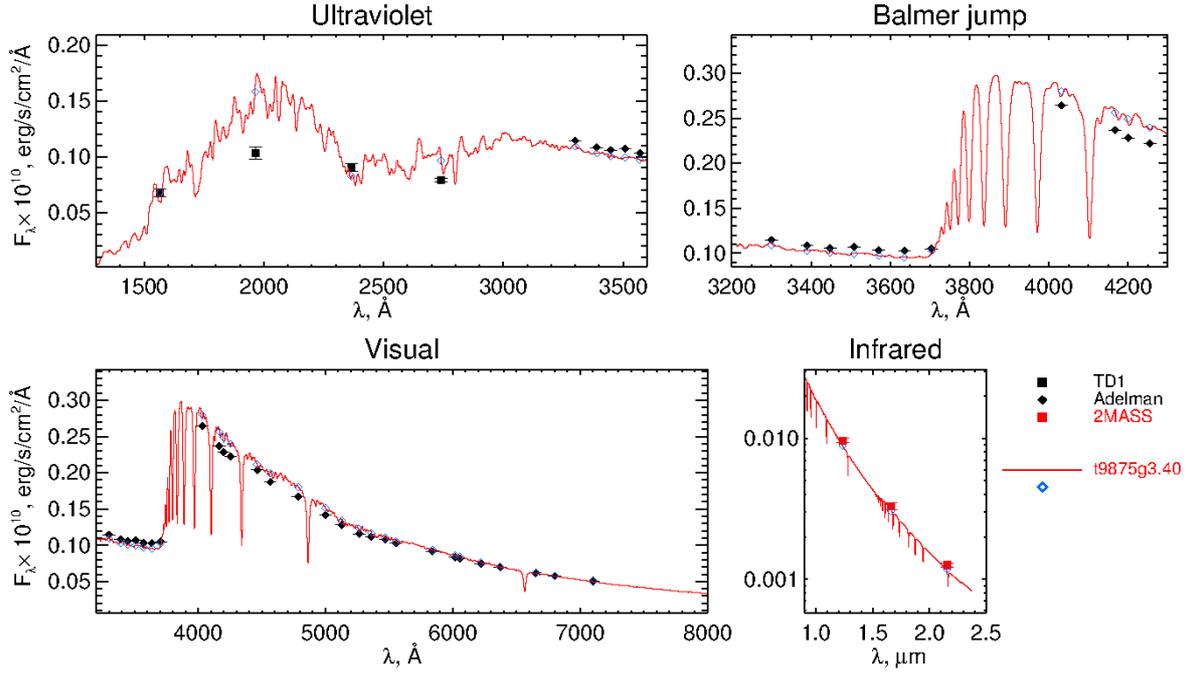}
	\caption{Same as in Fig.~\ref{sed-188041}, but for HD~111133.}
	\label{sed-111133}
\end{figure*}

We used parallax 5.21(7)~mas in our fitting procedure.
As for HD~188041, the model with solar helium abundance fits somewhat better
the observed energy distribution. Finally, we chose a model with \teff=9875~K, \lgg=3.4 derived from SED fit with E(B-V) taken from \citet{2005AJ....130..659A} as for two other stars. Comparison of the observed fluxes with
the theoretical ones is shown in Fig.~\ref{sed-111133}. Note that even after convergence 
we could not reach a good agreement between the observed
and theoretical energy distributions as in the case of HD~188041. 
This is especially noticeable in the UV spectral region up to the Balmer jump.
Ignoring TD1 fluxes we obtained best fit model that is about $100$~K hotter but could not constrain
stellar \lgg\ (not shown on the plot).

Using the largest parallax value of 6.23(93) \citep{1997ESASP1200.....E} we derive the same effective temperature and gravity as before, but
smaller radius $R/R_\odot$=2.92(44) (with the parallax uncertainty taken into account in the error estimate).

\subsubsection{HD~204411}
\begin{figure*}
	\includegraphics[width=0.9\textwidth]{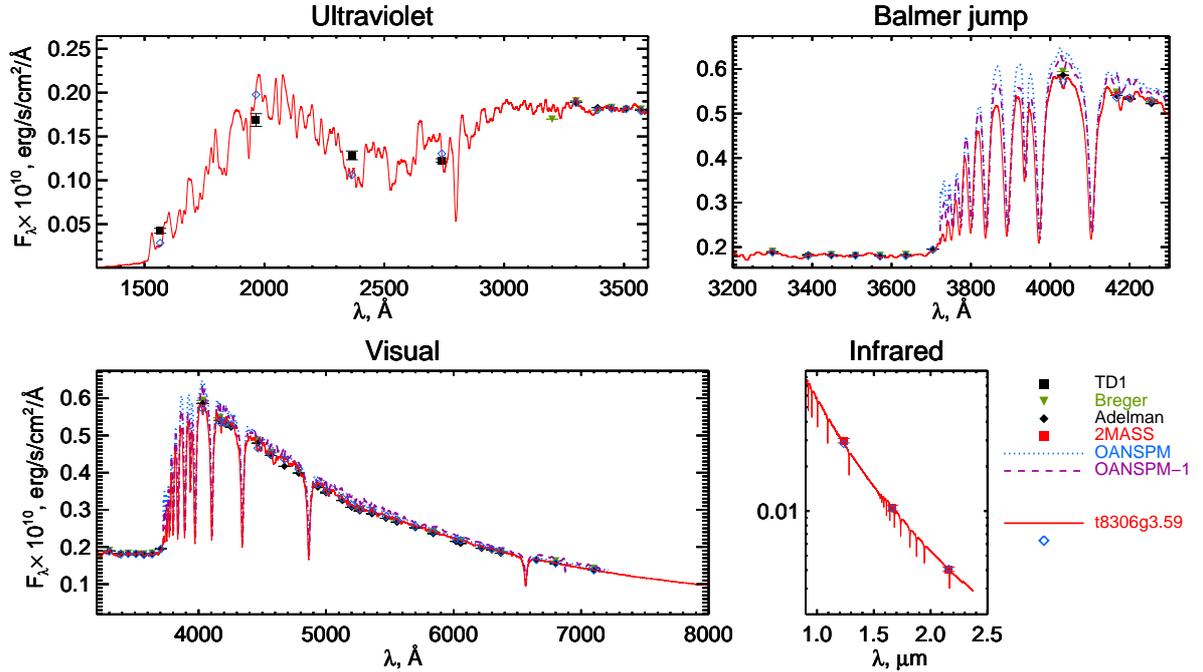}
	\caption{Same as in Fig.~\ref{sed-188041} but for HD~204411.
    We also show SEDs obtaiend with the Boller \& Chivens long-slit spectrograph (OAN SPM, M\'exico)
    and calibrated to absolute flux units using two different calibrators and marked as OANSPM (calibrator BD+33d2642) 
    and OANSPM-1 (calibrator HD~192281), respectively (see the figure legend).}
	\label{sed-204411}
\end{figure*}

\noindent
We compare the observed and theoretical fluxes for HD~204411 on Fig.~\ref{sed-204411}.
We obtained the best fit of the theoretical flux to the observed one after two iterations and for the model with 
\teff=8300(22)~K, \lgg=3.6(5), $R/R_\odot$=4.42(2) 
using parallax 8.33(12)~mas and low resolution spectro-photometry from \citet{1989A&AS...81..221A}. 
We obtain very similar result if we use spectro-photometry from \citet{1976ApJS...32....7B}:
\teff=8295(30)~K, \lgg=3.5(6), $R/R_\odot$=4.45(3). 
If we use observations obtained with the Boller~\&~Chivens spectrometer, we obtained very close estimates
of \teff=8356(11)~K, \lgg=3.41(4), $R/R_\odot$=4.45(1) and 
\teff=8351(10)~K, \lgg=3.23(4), $R/R_\odot$=4.46(1) for the two different calibrators BD+33d2642 and HD~192281, respectively.
However, in order to achieve such a good match we had to ignore fluxes blueward $4000$~\AA\, because of poor sensitivity
of the spectrometer at these short wavelengths, as can be seen from Fig.~\ref{sed-204411}.

\subsubsection{Adopted fundamental parameters}
\noindent
As was shown above, determination of fundamental parameters relies on datasets obtained with different instruments and missions
and for the same star there could be several alternative observations available. Also, our errors on the temperature, gravity, and radius 
listed above result from the $\chi^2$ minimization algorithm and are obviously underestimated because of uncertainties
in model atmopsheres, different data sources and accuracy of their calibration, etc.
Therefore, we decided to adopt concervative uncertainties
that account for the scatter in our estimates. Moreover, because for all our targets we always have data from at least three observing campangs
(TD1 for UV, \citet{1989A&AS...81..221A} for VIS, and 2MASS for IR), we adopted final parameters from the fit to these three data sets, adding the parallax uncertainty to radius estimates:
\teff=8770(150)~K, \lgg=4.2(1), $R/R_\odot$ = 2.39(5) for HD~188041, 
\teff=9770(200)~K, \lgg=4.0(2), $R/R_\odot$ = 3.49(7) for HD~111133, and
\teff=8300(150)~K, \lgg=3.6(1), $R/R_\odot$ = 4.42(7) for HD~224411, respectively.
The luminosity of stars were calculated from the derived values
of the effective temperature and radius.  The finally adopted parameters are listed in Table ~\ref{HRD-table}.
For HD~111133 we provide two sets of parameters based on Hipparcos and GAIA parallaxes, respectively.

Our model SED's clearly show a presence of 5200\AA\, depression usually observed in Ap stars.  From our theoretical
fluxes we computed $\Delta$a-index according to \citet{2007A&A...469.1083K} and found it to be 56 mmag for HD111133, 60 mmag for HD188041, and 20 mmag for HD204411. Theoretical $\Delta$a values agree very well with the observed data: 56 mmag (HD~111133), 60 mmag (HD~188041) and 19 mmag (HD~204411). For the first two stars the observed values are taken from \citet{1976A&A....51..223M}, while for HD~204411 $\Delta$a-index is taken from
\citet{1998A&AS..128..573M}. This comparison gives a credit to the results of our modelling.

\section{Interferometric observations}\label{interferometry}

\begin{figure*}
	\includegraphics[width=0.45\textwidth]{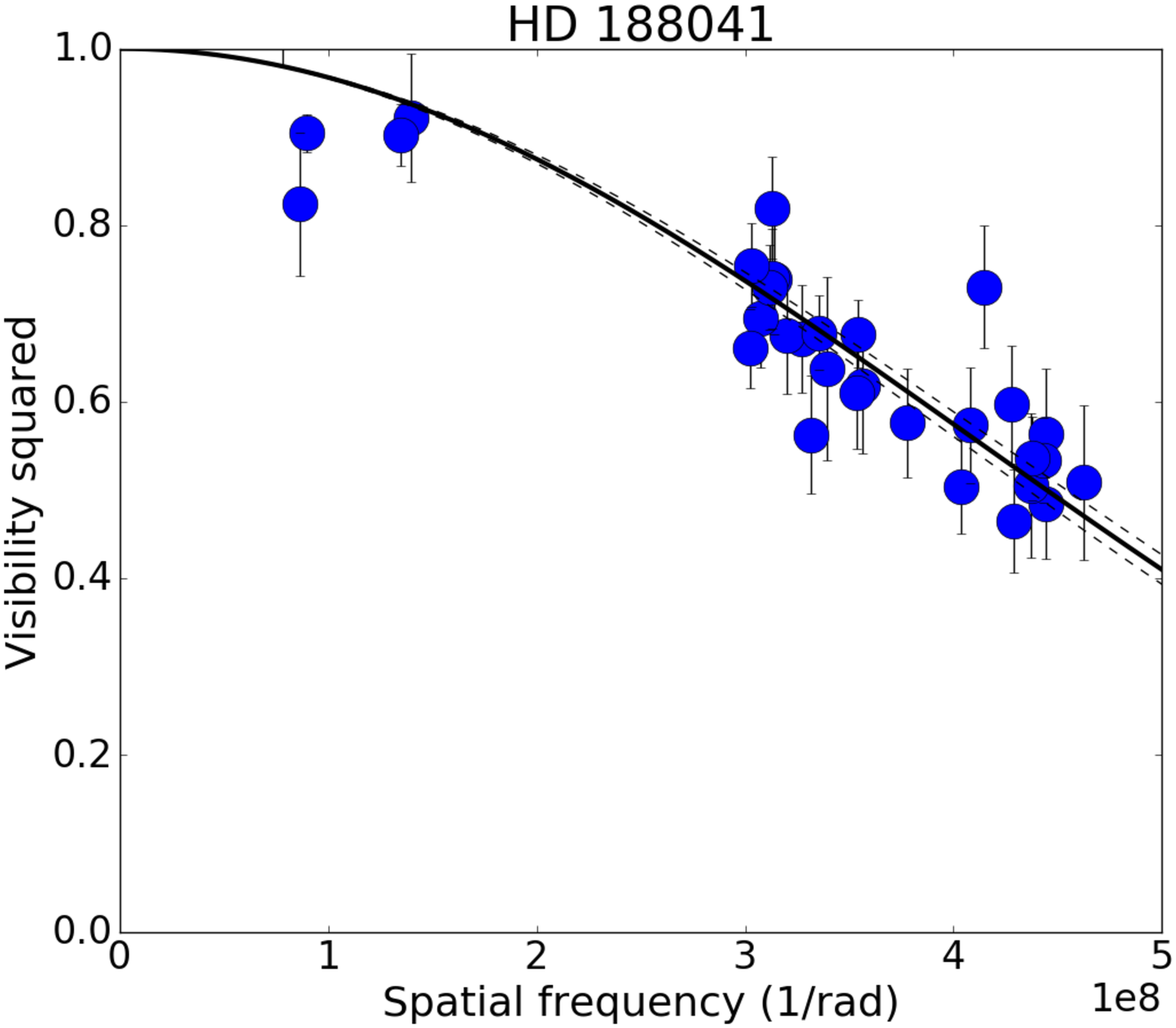}\hspace{10mm}\includegraphics[width=0.45\textwidth]{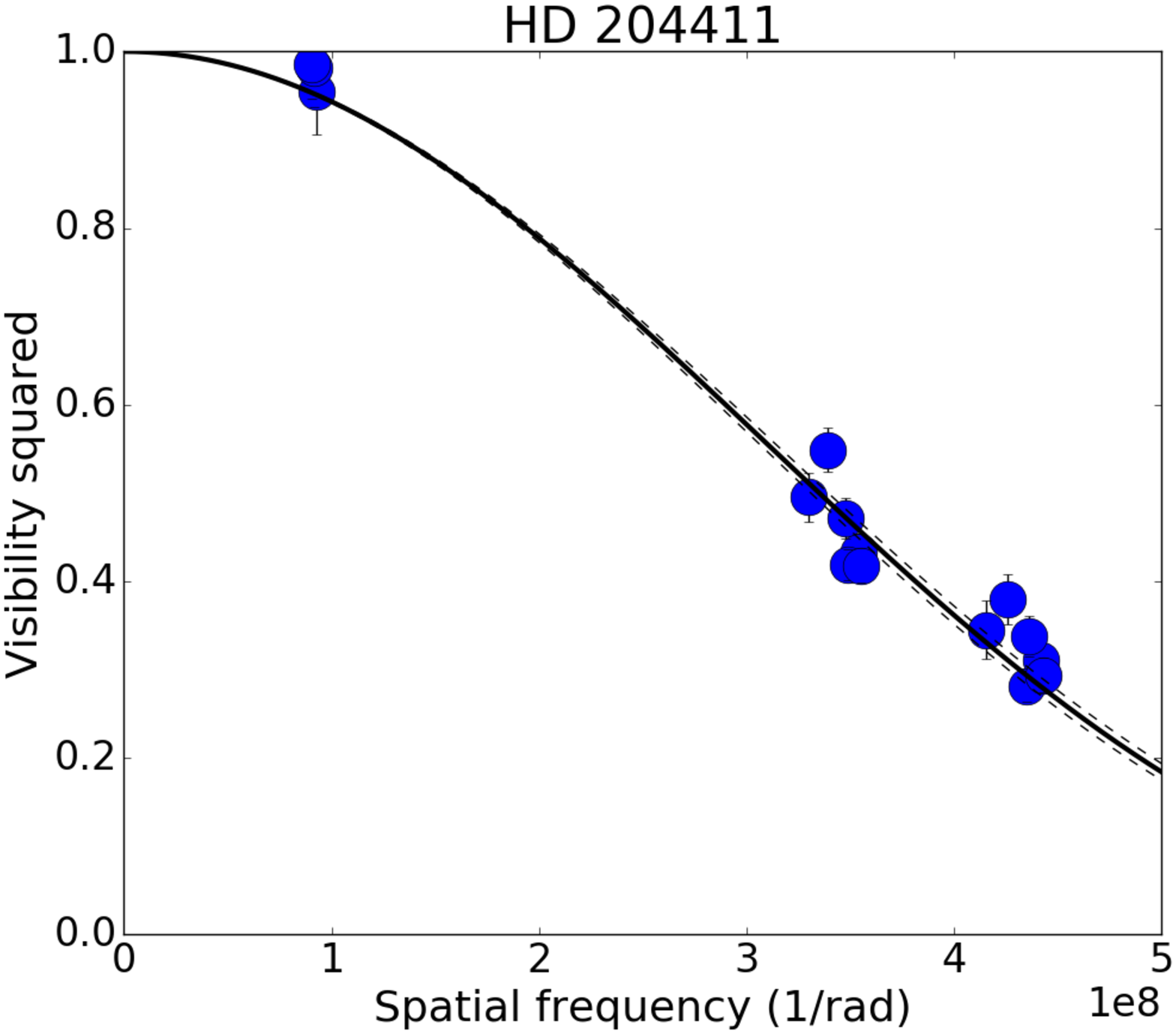}
	\caption{Squared visibility versus spatial frequency for HD~188041 (left panel) and for HD~204411 (right panel). 
    Solid lines represent the best uniform-disk models provided by LITpro.}
	\label{interf}
\end{figure*}

The two stars HD~188041 and HD~204411 were observed between June 2015 and June 2017 with the VEGA instrument \citep{vega} 
at the CHARA interferometric array \citep{chara} with the medium spectral resolution mode (R$\sim$6000). 
We recorded 15 datasets on HD~188041 and 6 datasets on HD~204411 with interferometric baselines spanning from about 50~m to 310~m. 
Each target observation of about 10 minutes is sandwiched with observations of reference stars to calibrate the instrumental transfer function. 
We selected calibrators bright and small enough, and close to the target thanks to the JMMC SearchCal service 
(\cite{searchcal}\footnote{\small www.jmmc.fr/searchcal}): HD~188293, HD~188294, HD~194244, and HD~203245. 
We used the standard VEGA data reduction pipeline \citep{vega} and the angular diameter of the reference stars provided 
by the JSDC2 catalogue \citep{JSDC2} to compute the calibrated squared visibility of each measurement. 
Fig.~\ref{interf} shows the results of interferometric measurements.

We used the model fitting tool LITpro\footnote{$\textrm{\small www.jmmc.fr/litpro\_page.htm}$} to determine the uniform-disk angular diameter 
of our targets, and the Claret tables \citep{2011yCat..35290075C} to determine the limb-darkened angular diameters using a linear limb-darkening 
law in the R band. For HD~188041, we obtained an angular diameter of $\theta_{UD}$~=~0.238(5)~mas, 
a limb-darkened angular diameter of $\theta_{UD}$~=~0.246(5)~mas considering an effective temperature ranging from 8750 to 9250~K 
and a surface gravity ranging from 4 to 4.5. We used parallax values from GAIA DR2 release \citep{2018A&A...616A...1G} with the error estimates as described in Section~\ref{flux}. Assuming a  parallax of 11.72(11), we deduced a radius of $R$~=~2.26(5)~$R_\odot$ for HD~188041. 

For HD~204411, we obtained an angular diameter of $\theta_{UD}$~=~0.316(4)~mas, a limb-darkened angular diameter of $\theta_{UD}$~=~0.328(4)~mas 
considering an effective temperature ranging from 8250 to 8750~K and a surface gravity ranging from 4 to 4.5. 
Assuming a parallax of 8.33(12), we deduced a radius of $R$~=~4.23(8)~$R_\odot$ for HD~204411.

\begin{figure*}
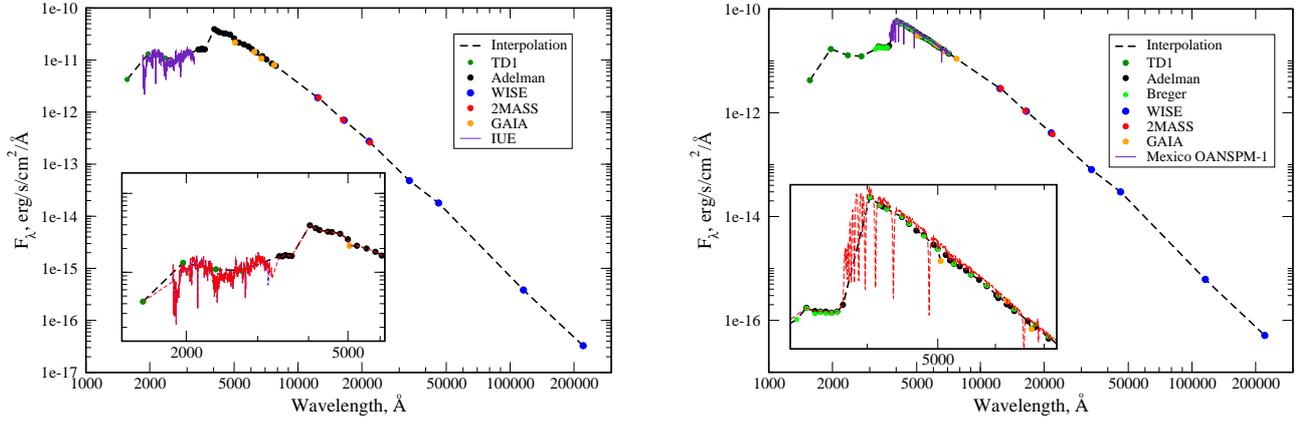

	\includegraphics[width=0.45\textwidth]{HD188041_flux_c.eps}\hspace{10mm}\includegraphics[width=0.45\textwidth]{HD204411_flux_c.eps}
	\caption{Observed flux distribution of HD~188041 (left panel) and  HD~204411(right column). Interpolation curves are shown by dashed lines. \textit{Insets} show parts of flux distributions with different overlapping observations.}
	\label{flux-distr}
\end{figure*}

The measured stellar radii allow us to derive effective temperatures from the bolometric flux using Stefan-Boltzmann law. 
The total flux for both stars was derived using photometric and spectrophotometric observations calibrated in absolute units. 
Photometric data were extracted from TD1 \citep{1978csuf.book.....T}, IUE archive\footnote{\tt http://archive.stsci.edu/iue/}, GAIA DR2 \citep{2018A&A...616A...1G}, 2MASS \citep{2003yCat.2246....0C}, 
WISE \citep{2014yCat.2328....0C} catalogues, while UV spectra and optical spectrophotometry were taken from  IUE archive\footnote{\tt http://archive.stsci.edu/iue/}, from \citet{1989A&AS...81..221A} and \citet{1976ApJS...32....7B} catalogues. For HD~204411 we also plotted M\'exico observations described in Section~\ref{sec:sed-obs}
Fig.~\ref{flux-distr} shows the observed flux distribution for HD~188041 and HD~204411. 
Flux integration was performed using the spline interpolation along the observed data points. 
The dashed line represents the resulting SED employed for integration. UV zero point was taken at 912~\AA. 
We obtained the bolometric fluxes $F=1.410\times10^{-7} erg/cm^2/s$ (with TD1) and $F=1.364\times10^{-7} erg/cm^2/s$ (with IUE) for HD~188041. In case of HD~204411 with different optical spectra/spectrophotometry we obtained $F=1.915\times10^{-7} erg/cm^2/s$ (Adelman/Breger) and  $F=2.000\times10^{-7} erg/cm^2/s$ (OANSPM-1)
Employing interferometric radii and assuming 10\% uncertainty in flux determination we deduced effective temperatures 
of our stars to be 9060/8990(250)~K (HD~188041)  and 8560/8650(230)~K (HD~204411). If we employ the radii derived by means of spectroscopy, 
the corresponding temperatures are 8880/8810 and 8360/8460~K, which agrees well with the spectroscopic determinations and validates
our flux integration procedure and the adopted effective temperature uncertainty. 

We also calculated bolometric flux $F=8.067\times10^{-8} erg/cm^2/s$ for HD~111133 and estimated an effective temperature as 9590(250)~K 
using spectroscopic radius. Again, the effective temperature derived from integrated flux agrees within the error with 9770(200) 
derived from self-consistent spectroscopic analysis.

\section{Position of Ap stars on the HR diagram}\label{evol}

Table~\ref{HRD-table} lists Ap stars with the fundamental parameters derived from the detailed self-consistent spectroscopic analysis 
taking into account abundance anomalies and stratification of the most important elements. 
We also added four more stars, HD~8441, HD~133792, HD~40312, and HD~112185 for which fundamental parameters 
were derived from detailed but not fully self-consistent spectroscopic analysis \citep{2012AstL...38..721T, 2006A&A...460..831K, 2019A&A...621A..47K}. 
These stars are marked by italic font in Table~\ref{HRD-table}. Note that the luminosities of the first two stars were recalculated 
based on the parallaxes from DR2 catalogue (3.44(10) and 5.49(5)~mas). 
The values of the surface magnetic fields for most stars are taken from \citet{2008A&A...480..811R} and references therein. 
For HD~40312 and HD~11218 the magnetic fields \bs\, are taken from Table~4 of \citet{2019A&A...621A..47K}. 
No estimates of the \bs\, in HD~8441 and HD~103498 are available, but according to \citet{2007A&A...475.1053A} 
their longitudinal fields vary between -200 and 200~G. Assuming a commonly found in other stars
poloidal dipole-dominant configuration, we find that most probably the surface magnetic field in these stars does not exceed 1~kG. 
The last four columns contain fundamental stellar parameters derived by interferometry together with the corresponding references. 

	\begin{table*}
		\caption{Fundamental parameters of Ap stars.}
	\scriptsize
		\begin{tabular}{r|cccccl|cccl}
			\hline 
			HD  &\multicolumn{6}{c}{Spectroscopy}&\multicolumn{4}{c}{Interferometry}\\
			& \teff &\lgg &$R/R_{\odot}$& $\log(L/L_{\odot})$& \bs,kG &Reference &$R/R_{\odot}$& \teff & $\log(L/L_{\odot})$&Reference\\
			\hline 
         {\it 8441}& ~9130(100) &3.45(17)&         & 2.21(12)&         &{\citet{2012AstL...38..721T}}&         &                  &                             \\   
			24712  & ~7250(100) &4.10(15)&1.77(04) & 0.89(07)&~~2.3    &{\citet{2009A&A...499..879S}}&1.77(06) &7235(280)&0.89(07)&{\citet{2016A&A...590A.117P}}\\  
        {\it 40312}& 10400(100) &3.6(1)  &4.64(17) & 2.35(06)&~~0.4    &{\citet{2019A&A...621A..47K}}&         &                  &                             \\   
			101065 & ~6400(200) &4.20(20)&1.98(03) & 0.77(06)&~~2.3    &{\citet{2010A&A...520A..88S}}& & & &\\                                      
			103498 & ~9500(150) &3.60(10)&4.39(75) & 2.15(16)&         &{\citet{2011MNRAS.417..444P}}& & & &                             \\  
			111133$^1$ & ~9875(200) &3.40(20)&3.44(07) & 2.00(04)&~~4.0    & Current work               & & & &\\                                       
            111133$^2$ & ~9875(200) &3.40(20)&2.92(44) & 1.84(17)&~~4.0    & Current work               & & & &\\                                       
       {\it 112185}& ~9200(200) &3.6(1)  &4.08(14) & 2.03(08)&~~0.1    &{\citet{2019A&A...621A..47K}}&         &                  &                             \\   
			128898 & ~7500(130) &4.10(15)&1.94(01) & 1.03(03)&~~2.0    &{\citet{2009A&A...499..851K}}&1.97(07) &7420(170)&1.02(02)&{\citet{2008MNRAS.386.2039B}}\\  
       {\it 133792}& ~9400(200) &3.7(1)~~~~&3.9(5)~~ & 2.02(10)&~~1.1  &{\citet{2006A&A...460..831K}}&         &                  &                             \\   
			137909 & ~8100(150) &4.00(15)&2.47(07) & 1.37(08)&~~5.4    &{\citet{2013A&A...551A..14S}}&2.63(09) &8160(200)&1.44(03)&{\citet{2010A&A...512A..55B}}\\  
			137949 & ~7400(150) &4.00(15)&2.13(13) & 1.09(15)&~~5.0    &{\citet{2013A&A...551A..14S}}& &  & &\\                                      
			176232 & ~7550(050) &3.80(10)&2.46(06) & 1.29(04)&~~1.5    &{\citet{2013A&A...552A..28N}}&2.32(09) &7800(170)&1.26(02)&{\citet{2013A&A...559A..21P}}\\  
			188041 & ~8770(150) &4.20(10)&2.39(05) & 1.48(03)&~~3.6    & Current work                &2.26(05) &9060(250)&1.49(04)&Current work\\
			201601 & ~7550(150) &4.00(10)&2.07(05) & 1.10(07)&~~4.0    &{\citet{2013A&A...551A..14S}}&2.20(12) &7364(250)&1.11(05)&{\citet{2011A&A...526A..89P}}\\  
			204411 & ~8300(150) &3.60(10)&4.42(07) & 1.92(04)&$\leq$0.8& Current work                &4.23(08) &8560(230)&1.96(04)&Current work\\                   
			\hline
            \multicolumn{8}{p{.6\textwidth}}{Fundamental parameters derived by using original Hipparcos parallax$^1$ and by GAIA DR2$^2$.
} \\		
     \end{tabular}
      \label{HRD-table}
	\end{table*}

\begin{figure}
	\centering
	\includegraphics[width=0.45\textwidth,clip]{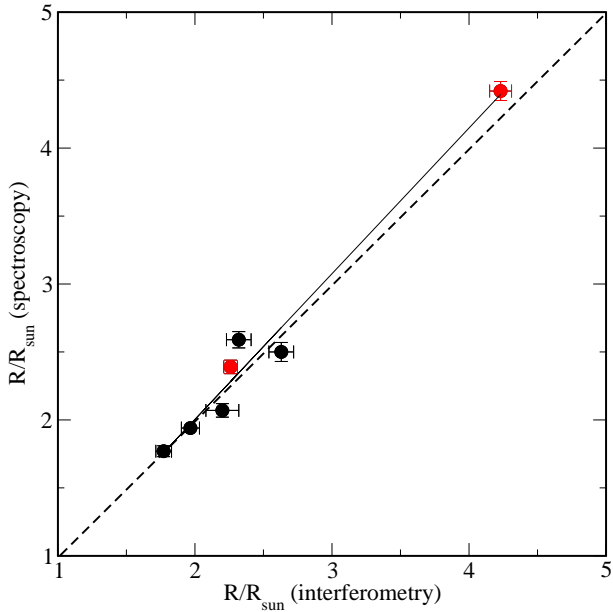}
	\caption{Comparison between stellar radii derived by means of spectroscopy and photometry. 
    Data from the present work is shown by filled red circles. Solid and dashed lines indicate regression line and line of equal values.}
	\label{Rad-comp}
\end{figure}

Comparison of the stellar radii derived by means of spectroscopy and interferometry (see Fig.~\ref{Rad-comp}) shows a reasonable agreement. 
Spectroscopic radii are larger by 7~\% on average, which is well within 2$\sigma$ of the interferometric measurements. 
It means that spectroscopically derived radii will provide us with rather accurate estimates of this fundamental parameter for fainter Ap stars 
where interferometric observations are not yet possible. 

\begin{figure}
	\centering
	\includegraphics[width=0.45\textwidth, clip]{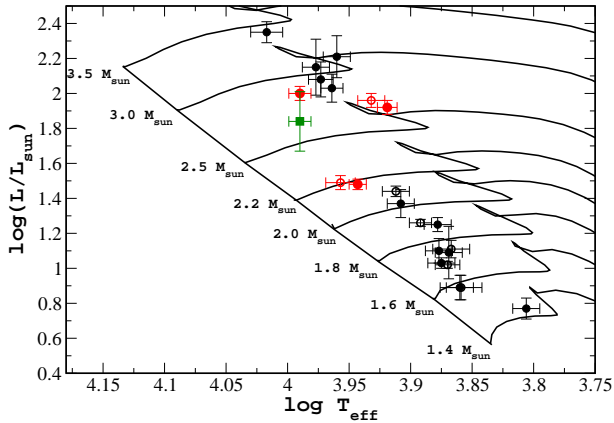}
	\caption{Ap stars on HR-diagram. Position of HD~111133, HD~188041, and HD~204411 from this work are shown by filled red circles (spectroscopy) 
    and by open red circles (interferometry). Position of other stars are indicated by black filled circles (spectroscopy) and by open black circles 
    (interferometry). Position of HD~111133 based on original Hipparcos parallax is marked by filled green square.}
	\label{HRD}
\end{figure}

Figure ~\ref{HRD} shows the position of stars from Table~\ref{HRD-table} 
on the H-R diagram.
Evolutionary tracks for the standard chemical composition models are taken from \citet{2000A&AS..141..371G}. 
Filled and open circles represent star's position based on spectroscopic and interferometric data, respectively. 
The position of HD~111133 with the luminosity, calculated using the effective temperature and radius 
based on parallax value from the original Hipparcos catalogue is shown by filled green square.

Although the number of objects shown in Fig. ~\ref{HRD} is not enough for robust statistical analysis, some preliminary conclusions may be pointed
out already now.
Our sample of stars is clearly divided in two groups. About half of stars are located in the middle part of MS, 
which agrees well with the results of evolutionary study of Ap stars by \citet{2006A&A...450..763K}. 
These stars possess moderately large magnetic fields and large REE overabundances. Group of  stars with small magnetic fields 
(HD~8441, HD~40312, HD~103498, HD~112185 HD~133792, HD~204411) are close to finish their MS lives. 
Some of these stars also have minimal anomalies of the rare earth elements. 
Our results on field Ap stars are in line with the results of other evolutionary studies in Ap stars. They confirm the conclusion 
that the magnetic field strength decreases with age \citep{2006A&A...450..763K}. They also support the conclusion drawn on the basis 
of an analysis of secular abundance variations in the atmospheres of Ap stars belonging to clusters of different ages that the REE anomaly
weakens with time \citep{2014A&A...561A.147B}. The only exception is HD~111133 which possess rather strong magnetic field 
but is located close to the group of evolved stars. As follows from Section~\ref{HD111133} SED fit for this star 
is not as good as for other stars, in particular, in UV region. 
We also note the large parallax difference in Hipparcos and GAIA catalogues. 
The use of parallax from the original Hipparcos catalogue gives better result because it brings HD~111133 
closer to group of stars with strong magnetic field, just as expected.
It will be important to perform interferometric measurements for this star to pin down its position on H-R diagram. 
In any case, we need to analyse more objects with the reliably defined fundamental parameters in order to constrain evolution history 
of Ap stars.

\section{Conclusion}

In this work we carried out a detailed atmospheric analysis of three Ap stars employing high-resolution spectroscopy
and (spectro)-photometry calibrated to flux units. 
For two of them, HD~188041 and HD~204411, we additionally obtained interferometric estimates of their angular diameters
that allowed us to derive model-independent stellar radii. For HD~204411 we also analized medium resolution spectroscopic 
observations obtained with the Boller \& Chivens long-slit spectrograph mounted at the $2.1$\,m telescope of the Observatorio
Astron\'omico Nacional at San Pedro M\'artir, Baja California,
M\'exico (OAN SPM). The results of our study are summarized as follows:
\begin{itemize}
\item
We find a very good agreement between effective temperatures and radii derived by means of spectroscopy (employing model atmospheres)
and interferometry (employing observed SED and radii derived from stellar angular diameters and parallaxes).
It means that spectroscopically derived radii will provide us with rather accurate estimates of fundamental parameters for fainter Ap stars 
for which interferometric observations are not yet possible.
\item
Our empirical abundance stratification analysis is in a good agreement with the predictions of modern diffusion models,
with the exception of Ca deficiency in the atmosphere of HD~111133. This needs additional investigation.
\item
Our results are consistent with the conclusion drawn in previous studies that the strength of the magnetic field and abundance anomalies
weaken as stars age.
\item
For HD~118041 and HD~204411 we improved analysis of atmospheric abundances 
by including accurate Zeeman splitting in the equivalent width calculation.
\item
Spectroscopic observations obtained with Boller \& Chivens spectrograph (OAN SPM, M\'exico) and their calibration agree well with
similar data sets. However, the poor instrument performance in UV region around Balmer jump limits our ability
to constrain accurate surface gravity and alternative observations should be used instead.
\end{itemize}

\section*{Acknowledgements}
This research has made use of the data from WISE, 2MASS and GAIA DR2 catalogues through the VizieR catalogue access tool, CDS (Strasbourg, France), 
and of the SearchCal and LITPRO services of the Jean-Marie Mariotti Center. The use of the VALD database is acknowledged. 
The authors warmly thank F. Morand for the support during the interferometric observations with the VEGA instrument. 
VEGA is supported by French programs for stellar physics and high angular resolution PNPS and ASHRA, 
by the Nice Observatory and the Lagrange Department. This work is based upon observations obtained with 
the Georgia State University Center for High Angular Resolution Astronomy Array at Mount Wilson Observatory. 
The CHARA Array is supported by the National Science Foundation under Grant No. AST-1211929, AST-1411654, AST-1636624, and AST-1715788. 
Institutional support has been provided from the GSU College of Arts and Sciences and the GSU Office of 
the Vice President for Research and Economic Development. KP acknowledges financial support from LabEx OSUG@2020 
(Investissements d'avenir - ANR10LABX56).
GV acknowledges the RFBR grant N18-29-21030 for support of his
participation in spectrophotometric analysis of the star HD~204411.
GG and GV acknowledge the Chilean fund CONICYT REDES 180136 for
financial support of their international collaboration.
Some of the data presented in this paper were obtained from the Mikulski Archive for Space Telescopes (MAST). 
STScI is operated by the Association of Universities for Research in Astronomy, Inc., under NASA contract NAS5-26555. 
Support for MAST for non-HST data is provided by the NASA Office of Space Science via grant NNX13AC07G and by other grants and contracts. 
We thank the anonymous reviewer for very valuable remarks which allow to improve some of the results. 
\bibliographystyle{mnras}
\bibliography{reference}

\label{lastpage}

\end{document}